# Comprehensive Optimization of Interferometric Diffusing Wave Spectroscopy (iDWS)

Mingjun Zhao, Leah Dickstein, Akshay S. Nadig, Wenjun Zhou, Santosh Aparanji, Hector Garcia Estrada, Shing-Jiuan Liu, Ting Zhou, Weijian Yang, Aaron Lord, Vivek J. Srinivasan

*Abstract*—**It has been shown that light speckle fluctuations provide a means for noninvasive measurements of cerebral blood flow index (CBFi). While conventional Diffuse Correlation Spectroscopy (DCS) provides marginal brain sensitivity for CBFi in adult humans, new techniques have recently emerged to improve diffuse light throughput and thus, brain sensitivity. Here we further optimize one such approach, interferometric diffusing wave spectroscopy (iDWS), with respect to number of independent channels, camera duty cycle and full well capacity, incident power, noise and artifact mitigation, and data processing. We build the system on a cart and define conditions for stable operation. We show pulsatile CBFi monitoring at 4-4.5 cm source-collector separation in adults with moderate pigmentation (Fitzpatrick 4). We also report preliminary clinical measurements in the Neuro Intensive Care Unit (Neuro ICU). These results push the boundaries of iDWS CBFi monitoring performance beyond previous reports.**

## I. INTRODUCTION

The non-invasive and continuous measurement of cerebral blood flow (CBF) is a major goal in biomedical optics. Investigations with Diffuse Correlation Spectroscopy (DCS) have revealed that light speckle fluctuations can non-invasively measure an optical blood flow index (BFi) [1] that correlates with quantitative blood flow [2-4]. DCS has stringent requirements: first, sufficient spatial coherence to maintain speckle contrast and second, sensitivity to highly attenuated light that samples the brain through the intact scalp and skull in adult humans. Typically single or few mode photon counting [5] was the best available approach to meet these requirements. To improve signal-to-noise ratio (SNR), multiple single photon counting DCS channels were pooled, with the utility of this strategy ultimately being constrained by cost [6]. Critically, in applications such as non-invasive intracranial pressure monitoring which need a fraction-of-a-second temporal resolution [7, 8], the DCS source-collector (S-C) separation on the adult human head is typically reduced to 2-2.5 cm, sacrificing brain sensitivity to improve photon counts [9, 10]. For non-invasive measurements in adult humans, this tradeoff between brain sensitivity and SNR is a critical roadblock [11].

Inspired by DCS, numerous techniques have recently emerged to improve SNR for deep cerebral BFi (CBFi) measurements [1, 12]. For instance, one recent study used a 512x512 single-photon avalanche diode (SPAD) array to retrieve pulsatile CBFi via the human forehead with a 0.125 s integration time, based on 3.3 microsecond sampling of the intensity autocorrelation at 4 cm S-C separation [13]. Interferometric methods have also shown promise by affording parallel detection with relatively inexpensive sensor arrays [14-19]. Interferometric Diffusing Wave Spectroscopy (iDWS) at 852 nm with a 512 pixel CMOS array provided pulsatile CBFi with a 0.1 s integration time, based on 3 microsecond sampling of the field autocorrelation ($G_1$) at 3.5-4 cm S-C separation [17]. Here we further optimize the 852 nm iDWS system in the latter study with respect to number of independent channels, camera duty cycle and full well capacity (FWC), noise and artifact mitigation, and data processing. In doing so, we define key governing parameters for iDWS system performance, identifying areas where the system can be improved. We validate predicted improvements against experimental results on various iterations of the system. We build the system on a cart and define conditions for stable operation. Finally, we demonstrate overall improvement by measuring pulsatile CBFi with a 0.125 s integration time, based on 6 microsecond sampling of the field autocorrelation at 4-4.5 cm S-C separation in adults with moderate pigmentation (Fitzpatrick 4 [20]). We also report preliminary clinical measurements in the Neuro Intensive Care

This work was submitted on 10/15/2024.

This work was supported by National Institutes of Health: EB029747, EB032840, EY031469, OT2OD038130, NS137832; Research to Prevent Blindness: Stein Innovation Award and unrestricted grant; National Natural Science Foundation of China: 62475247, 62105315. *(Corresponding author: V. Srinivasan,* e-mail: Vivek.Srinivasan@nyulangone.org*).*

Mingjun Zhao, Leah Dickstein, Akshay S. Nadig, Santosh Aparanji, Hector Garcia Estrada, Ting Zhou, Aaron Lord and Vivek J. Srinivasan are with NYU Langone Health, New York, New York 10010, USA.

Wenjun Zhou is with China Jiliang University, Hangzhou, Zhejiang 310018, China, and University of California Davis, Davis, California 95616, USA.

Shing-Jiuan Liu and Weijian Yang are with University of California Davis, Davis, California 95616, USA.

Vivek Srinivasan and Wenjun Zhou are inventors on a related patent owned by UC Davis.

Unit (Neuro ICU). As the system uses a non-scientific grade CMOS array, it offers a favorable performance-to-cost ratio compared to current state-of-the-art SPAD arrays.

## II. SYSTEM PARAMETERS AND MEASUREMENTS

Our iDWS system (**Fig. 1**) boosts the weak optical field returning from the brain by coherent amplification with a stronger reference field. This principle, when applied to a scheme with a collection multimode fiber (MMF) and non-scientific CMOS sensor, enables highly parallelized detection with hundreds of channels near the shot noise limit [14, 17].

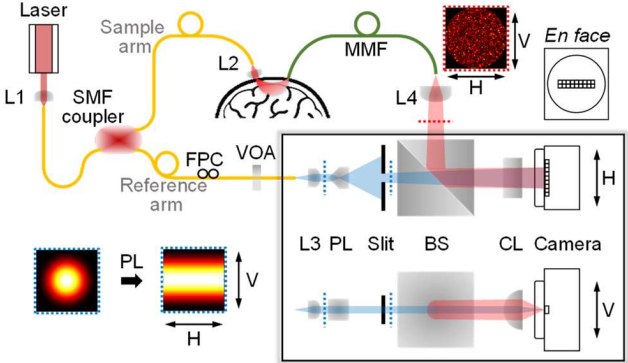

**Fig. 1** Diagram of iDWS system. H: horizontal view; V: vertical view; SMF: single-mode fiber; MMF: multimode fiber; L1 to L4, lenses; VOA, variable fiber-optic attenuator; PL, Powell lens; BS, beamsplitter; CL, cylindrical lens; FPC: fiber polarization controller. [17, 21]

### System parameters

#### Generalized autocorrelation function $G_1$

Temporal field correlations are of primary importance in recovering CBFi in iDWS. Spatial correlations between pixels, on the other hand, can determine system performance. To describe these correlations in a unified way, we introduce a generalized spatiotemporal autocorrelation function, $G_1(\chi, \tau)$, where $\chi$ is the spatial lag, and $\tau$ is the time lag. Since $G_1$ is sampled spatially and temporally by the sensor, we alternatively use $G_1(m, n)$, with a spatial lag index, $m = \chi/\Delta\chi$, where $\Delta\chi$ is the pixel pitch, and temporal lag index $n = \tau/\Delta\tau$, where $\Delta\tau$ is the temporal sampling period. We estimate $G_1(m, n)$ as

$$G_1(m,n) = \sum_{i=n+1}^{N} \sum_{p-q=m} s(p,i)s(q,i-n), \quad (1)$$

where $s(p,i)$ is the heterodyne signal in pixel $p$ at temporal lag index $i$, including shot noise. $N$ is the temporal window length.

#### Binning

A binning, or coherent pixel addition, procedure has been shown to improve autocorrelation measurements [17]. The binned autocorrelation, $G_{1,binned}(n)$, is given by

$$G_{1,binned}(n) = \sum_{i=n+1}^{N} \sum_{p,q} H_{p,q} s(p,i)s(q,i-n). \quad (2)$$

Note that binning includes spatial self-multiplications terms, where $p=q$, and cross-multiplication terms, where $p \neq q$. We have included a general binning matrix $H_{p,q}$ [17], which specifies the weighting of various terms in the sum. Note that **Eq. (2)** with $H_{p,q} = \delta_{p,q}$ (Kronecker delta function) corresponds to no binning, yielding $G_1(0, n)$ in **Eq. (1)**. Determination of the optimal binning matrix in the shot noise limit has been previously discussed [17].

#### Signal-to-additive-noise ratio (SANR)

The signal-to-additive-noise ratio (SANR) of the binned iDWS signal is a fundamental quantity, which is calculated from the autocorrelation estimate $G_{1,binned}(n)$ (**Fig. 2**a-b) [14]. The numerator is the mean squared heterodyne signal. The denominator is the additive white (uncorrelated after one lag) noise variance. The additive noise (**Fig. 2**b) is typically dominated by shot noise from the reference arm for a good iDWS system design [17]. In this shot noise limit, SANR is proportional to the product of the number of sample photons and the mutual coherence degree squared [14]. Note that the expression for SANR is inclusive of any binning.

#### Autocorrelation signal-to-noise ratio (SNRac):

Autocorrelation signal-to-noise ratio (SNRac) is the lag-wise estimate of the normalized SNR of the unbiased estimate of the autocorrelation across a temporal window of N samples (**Fig. 2**a). More generally, with binning, we have:

$$SNR_{ac,experiment}(n) = \frac{1}{var[g_{1,binned}(n)]}. \quad (3)$$

The SNRac is defined as the reciprocal of the variance across realizations or instances of the normalized autocorrelation estimate at a lag of n≠0 (**Eq. (3)**, **Fig. 2**c and d). Note that $SNR_{ac}$ does not completely determine the SNR of recovered decay rate or BFi, and should not be used to compare across different sampling rates.

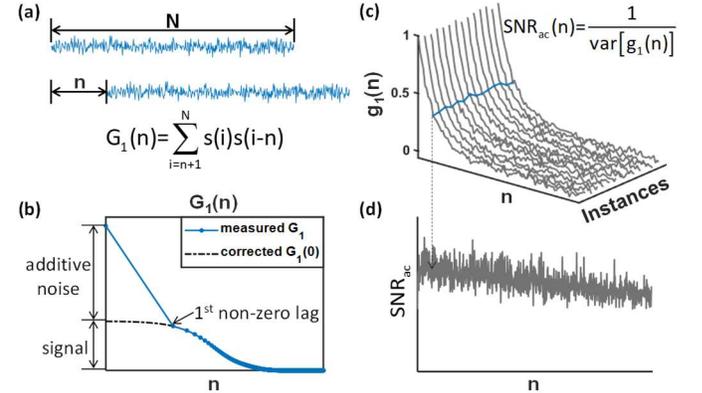

**Fig. 2** Definition of key iDWS signal-to-noise parameters. (a) Autocorrelation ($G_1$) estimation in iDWS signal of a single pixel. b) Signal-to-additive-noise-ratio (SANR), a fundamental property of the raw iDWS signal, can be determined from $G_1$. (c-d) SNRac estimation (**Eq. (3)**) from $g_1$, with removal of additive noise bias at zero lag. n: discrete lag index. N: temporal window length.

If the additive noise is Gaussian, the theoretical $SNR_{ac}$ is given by [17]: (**Fig. 2**a-b) (**Fig. 2**a-b)

$$SNR_{ac,theory}(n)=N_{channel}\times SANR^2\times(N-n). \quad (4)$$

Comparing the experimental and theoretical $SNR_{ac}$ is often a useful diagnostic.

### *Number of independent channels ($N_{channel}$)*

The number of independent channels ($N_{channel}$) depends on the spatial correlation of the heterodyne signals between pixels [17]. The maximal achievable $N_{channel}$ is the number of pixels for fully independent detectors (no spatial correlation between pixels, while the minimum (non-trivial) number of channels is 1 for identical signals measured across pixels).

### **System measurements**

#### *Full*

In a standard iDWS configuration, the full measurement includes the desired sample field autocorrelation (**Fig. 3**a), as well as undesirable contributions, which can be characterized by the measurements below.

#### *Source (S) out, collector (C) in reference*

By removing the source (S) from the sample, The S out C in measurement characterizes spurious paths (**Fig. 3**b) that contribute to artifactual autocorrelation decay, as well as contributions from the measurements below. The difference between this and S out C out reference yields the spurious path artifact. If spurious paths are eliminated, the S out C in reference $G_1$ should reduce to the S out C out reference $G_1$.

#### *Source (S) out, collector (C) out reference*

By blocking the source and collector from the sample (or blocking the sample beam before the beam combiner), the S out C out measurement characterizes autocorrelation of reference light caused by (for instance) environmental vibrations (**Fig. 3**c) as well as contributions from the measurement below. The difference between this and the dark $G_1$ yields system stability.

#### *Dark*

By blocking laser light from reaching the sensor, a dark measurement allows characterizing camera noise or oscillations independently (**Fig. 3**d).

### METHODS AND RESULTS

#### *Increasing independent channel count*

*Rationale:* As expressed in **Eq. (4)**, a higher independent channel count ($N_{channel}$) improves $SNR_{ac}$. This allows larger S-C separation to improve the depth sensitivity of CBFi [12].

*Experimental Approach:* Increasing MMF numerical aperture (NA) and core diameter (Φ) will increase the number of guided modes, and thus $N_{channel}$ for iDWS. To test this, we upgraded the iDWS system by replacing the previous 400 μm core, 0.22 NA MMF [sample arm collimator (L4 in **Fig. 1**): Thorlabs A220TM-B, NA = 0.26] with a 600 μm core, 0.37 NA MMF (sample arm collimator: Thorlabs A240TM-B, NA = 0.5).

*Improvements:* To validate the increase in $N_{channel}$, we measured the spatial correlation at zero temporal lag, $G_1$(m, n=0), with the previous 400 μm MMF and the new 600 μm MMF. **Fig. 4** shows a narrower spatial correlation with the 600 μm MMF (HWHM = 1.05, **Fig. 4**b) than with the 400 μm MMF (HWHM = 1.77, **Fig. 4**a). The corresponding $N_{channel}$ has been improved from 192 in prior work [17] to up to ~338 in this work.

As further validation of the effect of the number of modes on the spatial correlation, we simply switched the source wavelength from λ = 852 nm to 785 nm. As shown in **Fig. 4**c the spatial correlation is narrowed at the shorter wavelength with more guided modes (number of modes ∝1/ λ²).

To determine improvement in $SNR_{ac}$ by increasing $N_{channel}$, we acquired data from a phantom and from *in vivo* tissue, before and after the MMF upgrade. The phantom was intralipid diluted in distilled water, with theoretical values of $\mu_s'$ = 7 cm⁻¹ and $\mu_a$ = 0.045 cm⁻¹. Phantoms were measured at 3 cm S-C separation for 2 seconds. *In vivo* measurements were performed on human forehead, at 3 cm S-C separation, for 10 seconds. [All *in vivo* experimental procedures and protocols involved in this study were reviewed and approved by New York University Langone Institutional Review Board (NYU Langone IRB). Informed consent was obtained for all subjects involved.] We calculated $G_1$ with a 0.01 second integration time, corresponding to an integration window of N = 3334 (**Fig. 2**a) at Δτ = 3 μs. This choice of acquisition parameters yielded 200 instances of $G_1$ for the phantom and 1000 instances of $G_1$ *in vivo*.

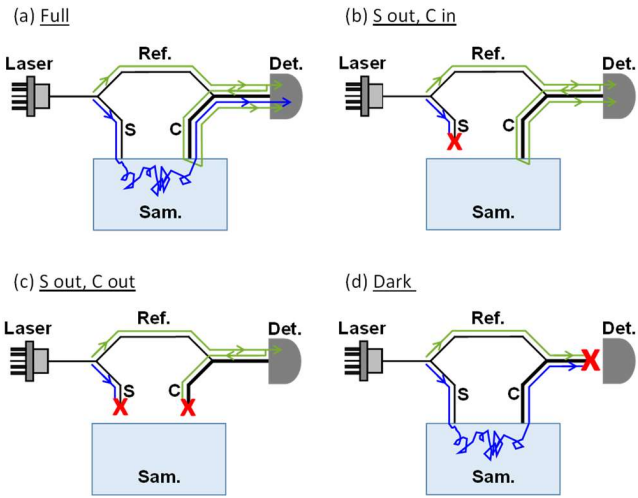

**Fig. 3** Measurement configurations for system characterization. S: source; C: collector; Ref.: reference arm; Sam.: sample; Det.: detector. A red X indicates that measures have been taken to ensure no light traverses the indicated path.

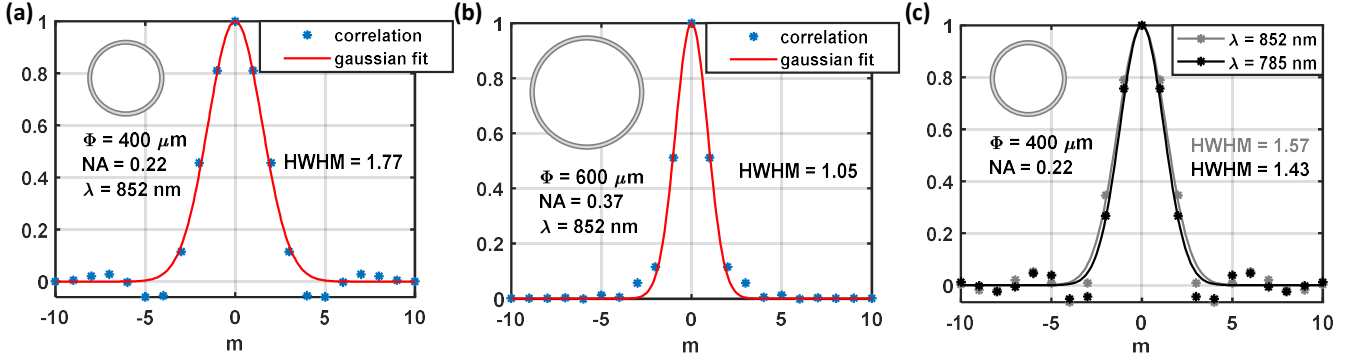

**Fig. 4** Spatial correlation narrows with increasing multimode fiber core size and NA (a-b), and decreasing wavelength (c). Both effects lead to more modes and therefore, independent channels.

Though the predicted decrease in SNR$_{ac}$ with n was observed (**Fig. 5**), the SNR$_{ac}$ measured experimentally [**Eq. (3)**] consistently underestimated the theoretical prediction [**Eq. (4)**], especially for the higher SANR phantom (**Fig. 5**a-b, **Table 1**). However, experiment approached theory at lower SANR levels found in *in vivo* (**Fig. 5**c-d, **Table 1**), excepting early lags where pulsatility increased the autocorrelation variability. The discrepancy at high SANRs shows that the assumptions of the additive noise limited expression [**Eq. (4)**] are invalid. One possible reason could be contributions to autocorrelation noise from speckle noise, which is not accounted for in **Eq. (4)**. Noise contributions from speckle are expected to be relatively more important if SANR is higher or if decorrelation rate is lower [22].

For both phantom (**Fig. 5**a-b) and *in vivo* (**Fig. 5**c-d) measurements, SNR$_{ac}$ was improved with the higher capacity ($\Phi$=600 μm, NA=0.37) MMF. Importantly, as seen in insets of **Fig. 5**c-d, the relative change of CBFi (rCBFi), obtained by fitting *in vivo* data (0.05 s integration time, 100 Hz sampling rate), also improved with higher SNR$_{ac}$ afforded by larger N$_{channel}$ (**Fig. 5**c-d).

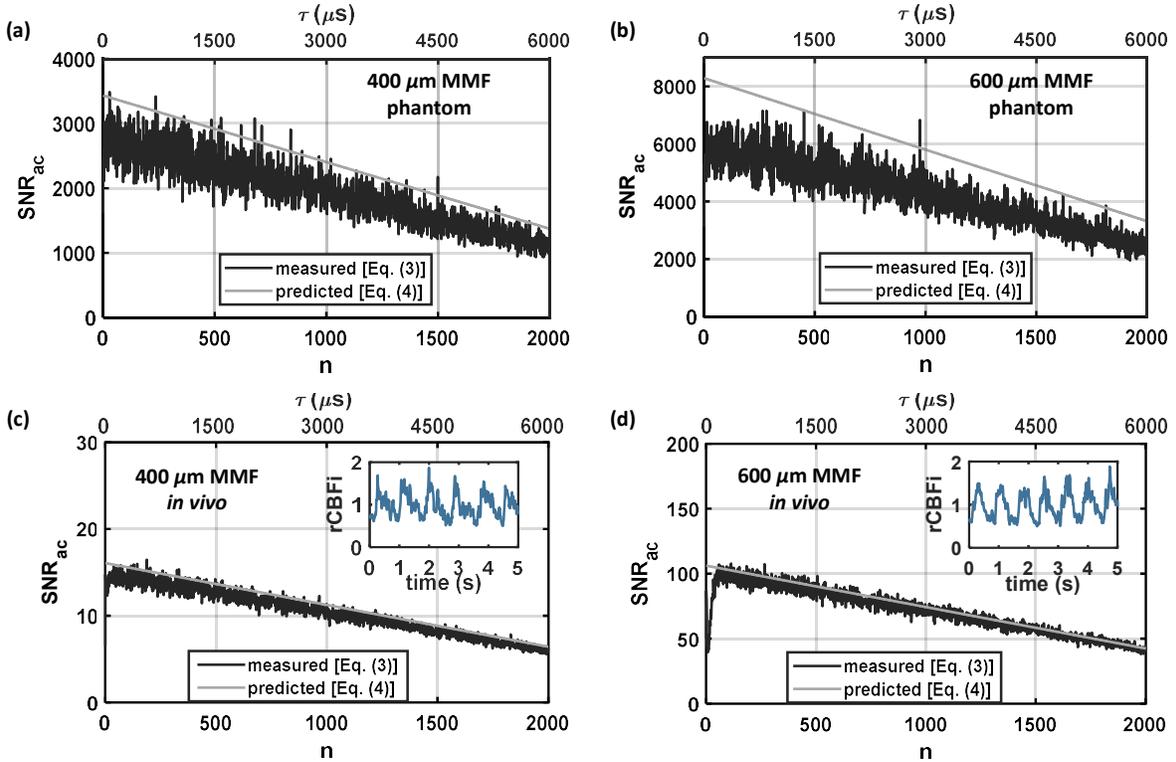

**Fig. 5** Comparison of measured SNR$_{ac}$ [**Eq. (3)**] and predicted SNR$_{ac}$ [**Eq. (4)**]. Note that discrepancy between measured and predicted SNR$_{ac}$ for the higher SNR$_{ac}$ phantom measurements at 3 cm S-C separation (a-b), and excellent agreement for the lower SNR$_{ac}$ in vivo measurements at 3 cm S-C separation (c-d). The MMF with bigger core and larger NA increases SNR$_{ac}$, leading to improved CBFi (insets). The higher variability in the measured SNR$_{ac}$ [**Eq. (3)**] of the phantom comparing to that of *in vivo* is due to the difference in the number of g$_1$ instances (**Fig. 2**a).

**Table 1.** Comparison of iDWS with two different multimode fibers. Core diameter (Φ), numerical aperture (NA), spatial correlation half-width-at-half maximum (HWHM), number of independent channels ($N_{channel}$), illumination power, signal-to-additive noise ratio (SANR), and ratio of experimental and theoretical autocorrelation signal-to-noise-ration ($SNR_{ac}$) [see **Eq. (3)-Eq. (4)**].

| | Φ (μm) | NA | HWHM | $N_{channel}$ | Illumination power (mW) | SANR | $\frac{SNR_{ac,experiment}}{SNR_{ac,theory}}$ |
|---|---|---|---|---|---|---|---|
| Phantom | 400 | 0.22 | 1.69 | 200.82 | 49.4 | 0.0716 | 0.80 |
| | 600 | 0.37 | 1.21 | 285.56 | 57.4 | 0.0933 | 0.73 |
| *in vivo* | 400 | 0.22 | 1.63 | 209.36 | ~60 | 0.0048 | 0.93 |
| | 600 | 0.37 | 1.06 | 338.48 | ~60 | 0.0097 | 0.97 |

*SANR was calculated with optimal pixel binning.

### Coherent filtering of spurious paths

*Rationale:* A very small portion of the light from reference arm (on the order of 0.1% of the reference power for 600 μm MMF, and on the order of 0.01% for 400 μm MMF) is backreflected by the camera or the unused port of the beam splitter (BS in **Fig. 1**), enters the sample via the collector fiber, then scatters into the collector along with light that has traversed the desired path from the source to collector (**Fig. 3b** and **Fig. 6a**). We call these 'spurious paths' as they do not traverse the sample in the desired way. Spurious paths are particularly problematic at long S-C separation where the light traversing the desired sample path is weak, and they are amplified by the strong reference field.

The power of light from the spurious paths is higher if either

1. The MMF in sample arm picks up more backreflected reference light. This applies when using larger/more core(s) and higher NA MMFs (see **Increasing independent channel count**).
2. Reference power is increased, as is required for a higher camera full well capacity (FWC), which requires higher reference power to fill the full well to the same gray level as a lower FWC.

Both situations call for a method to mitigate the signal from the spurious paths, without sacrificing the desired sample signal.

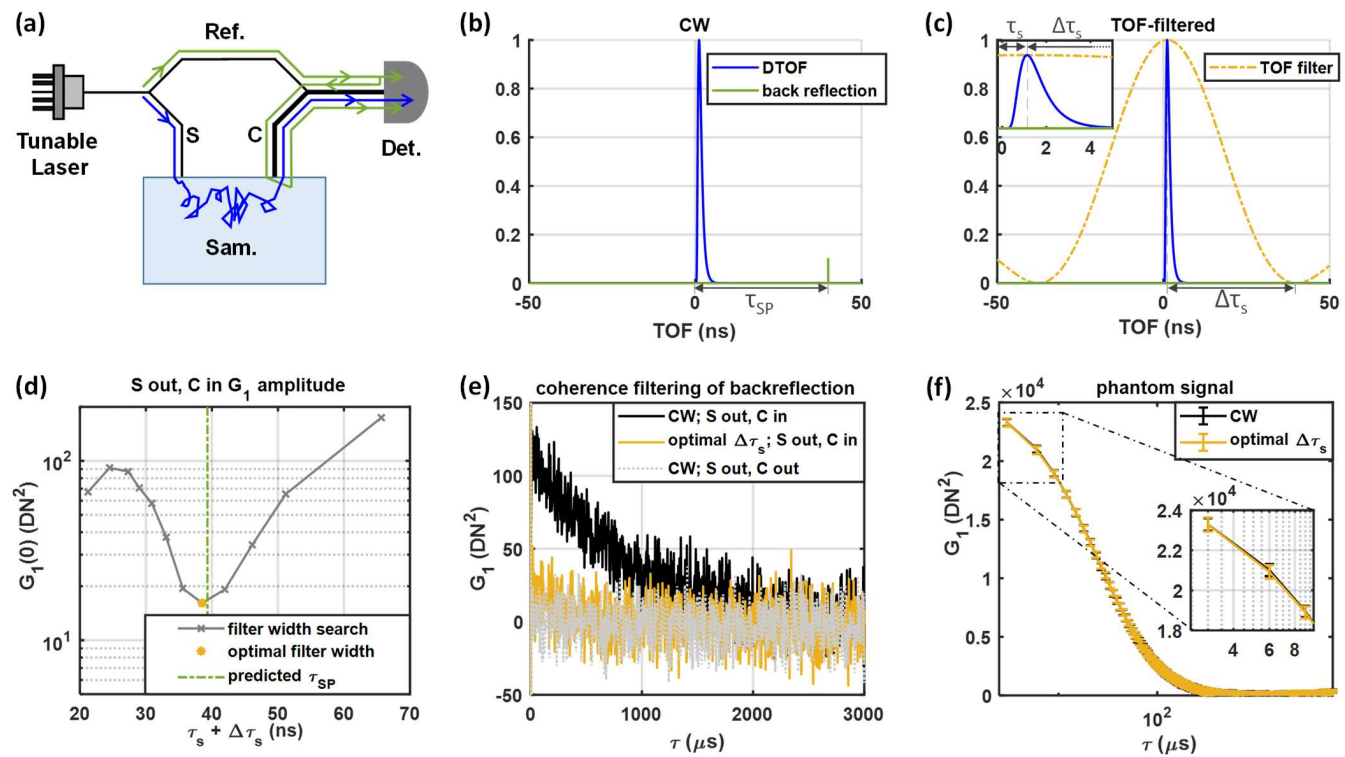

**Fig. 6** Coherence filtering removes $G_1$ contamination by spurious paths. (a) Full measurement setup, with spurious paths marked in green. (b) Artifacts in sample time-of-flight (TOF) domain caused by the spurious paths. (c) Coherence filter mitigates the artifacts caused by the spurious paths. Note that $\tau_s$ is too small to show in the main panel and is therefore only shown in the inset. (d) Optimizing coherence filer for spurious paths removal. (e) Optimal coherence filter removes artifacts caused by the spurious paths. (f) Optimal coherence filter does not significantly attenuate sample signal in a liquid phantom.

*Experimental Approach:* First, we included a fiber polarization controller in the reference arm to minimize reflections that are polarization-sensitive [21]. Second, we implemented a TOF filtering approach to attenuate the contribution of spurious paths to $G_1$ without additional post processing (**Fig. 6**c) [21]. Briefly, we sinusoidally tune the instantaneously narrowband laser rapidly in optical frequency during the sensor exposure time to decrease the effective temporal coherence [21]. Previously we used such a TOF filter to discriminate late sample TOFs [21]. Here, we adapt the approach to explicitly remove spurious paths while preserving the continuous wave (CW) sample signal.

For sinusoidal tuning, neglecting dispersion, the TOF filter, *H*, is given by

$$H\left(\tau_s' - \tau_s\right) = \left\{ J_0 \left[ \frac{2.40}{\Delta\tau_s} \left(\tau_s' - \tau_s\right) \right] \right\}^2, \quad (5)$$

where $\tau_s'$ is the TOF in the sample, $\tau_s$ is the center of the filter along TOF (**Fig. 6**c), given by the mismatch between reference and sample arms for zero TOF in the sample, $\Delta\tau_s$ is the effective temporal coherence (**Fig. 6**c), and $J_0$ is a Bessel function of the first kind. The key parameters of the TOF filter, $\Delta\tau_s$ and $\tau_s$ [21], can be varied via tuning range and interferometer path length mismatch, respectively.

To preserve the DTOF, we should choose the sample-reference path length difference to position $\tau_s$ close to the peak of sample DTOF (**Fig. 6**c). To eliminate spurious path artifacts, we should select $\Delta\tau_s$ so that the first minimum of the coherence filter in **Eq. (5)** attenuates $\tau_{SP}$, the TOF corresponding to the spurious paths (**Fig. 6**b-c). Therefore the optimal coherence filter should satisfy $\tau_{SP} = \tau_s + \Delta\tau_s$.

First, we experimentally estimated $\tau_{SP}$, the target TOF to be attenuated by the coherence filter. $\tau_{SP}$ is the difference between the total sample arm TOF except for the path in sample (**Fig. 6**a blue path) and the total spurious path TOF (**Fig. 6**a green path). For the instrument used in this study, $\tau_{SP}$ was measured to be 39.32 ns.

Next, we estimated the peak of the sample DTOF as $\tau_s = 0.67$ ns, by simulating the sample DTOF using a homogeneous semi-infinite diffusion model ($\mu_s' = 8$ cm$^{-1}$ and $\mu_a = 0.1$ cm$^{-1}$, 3 cm S-C separation) [23]. We then adjusted reference arm length with available patch cords to approximate the target $\tau_s = 0.67$ ns, and verified that we achieved $\tau_s = 0.49$ ns using the method reported in [21]. This 0.18 ns mismatch in $\tau_s$ is small compared to the width of the TOF filter, which must cover two passes through our 4 m sample collection fiber. This reasoning is confirmed by the validation tests described later in this section and **Fig. 6**f, which shows insignificant filtering of the DTOF.

Finally, to determine the optimal $\Delta\tau_s$, we varied $\Delta\tau_s$ such that $\tau_s + \Delta\tau_s$ varied around the expected $\tau_{SP} = 39.32$ ns (**Fig. 6**d). We found that $\Delta\tau_s = 38.02$ ns minimized the S out C in reference $G_1$ amplitude (**Fig. 6**d). Given possible inaccuracy in $\tau_{SP}$ estimation and limited resolution of experimental $\Delta\tau_s$, the experimentally determined optimal $\tau_s + \Delta\tau_s = 38.51$ ns (**Fig. 6**d 'star') aligns well with the predicted $\tau_{SP} = 39.32$ ns (**Fig. 6**d dashed vertical line).

With the optimal coherence filter, the S out C in reference $G_1$ was decreased nearly to CW S out C out reference $G_1$ (**Fig. 6**e). Also, the optimal coherence filter minimally attenuates the desired sample $G_1$ (**Fig. 6**f) [SANR at CW and optimal $\Delta\tau_s$ are $0.1153 \pm 0.0017$ and $0.1150 \pm 0.0017$ respectively; 1/e decay time at CW and optimal $\Delta\tau_s$ are $32.85 \pm 0.65$ μs and $32.90 \pm 0.65$ μs respectively]. If the sample signal had shown a significant TOF filtering effect (decreased $G_1$ SANR and/or altered decay rate), a more accurate choice of $\tau_s$ would have been required.

### Camera oscillation removal

*Rationale:* Camera pixels exhibit a correlated noise pattern (quasi-periodic, in-phase across pixels) in count levels (**Fig. 7**a-c). The frequency varies slightly over time, making it hard to remove using fixed pattern noise post processing techniques.

*Experimental Approach:* We added a spatial mean subtraction for each half of the sensor, prior to pixel binning, in our data processing pipeline [17]. A spatial mean subtraction will remove signal that are spatially correlated across pixels. Therefore, since the camera noise is highly correlated across pixels, in contrast to the short heterodyne signal spatial correlation length (see ***Increasing independent channel count***), spatial mean subtraction removes the quasi-periodic camera noise (**Fig. 7**d) while only modestly affecting the iDWS signal (**Fig. 7**e). This advance is essential to enable calculating and saving $G_1$ in real time.

Note that the spatial mean subtraction must be done for a subset of pixels that exhibit a correlated noise pattern (e.g. left and right halves of the centered 512 pixels for Basler spL4096-140km in this work, **Fig. 7**b-c). It should also be applied before pixel binning in case noise is not correlated over all pixels. Otherwise it will introduce the noise to adjacent pixels at the boundary of each region with a distinct noise pattern. Finally, we find that this method can remove any spatially correlated noise, introduced by the laser or other sources.

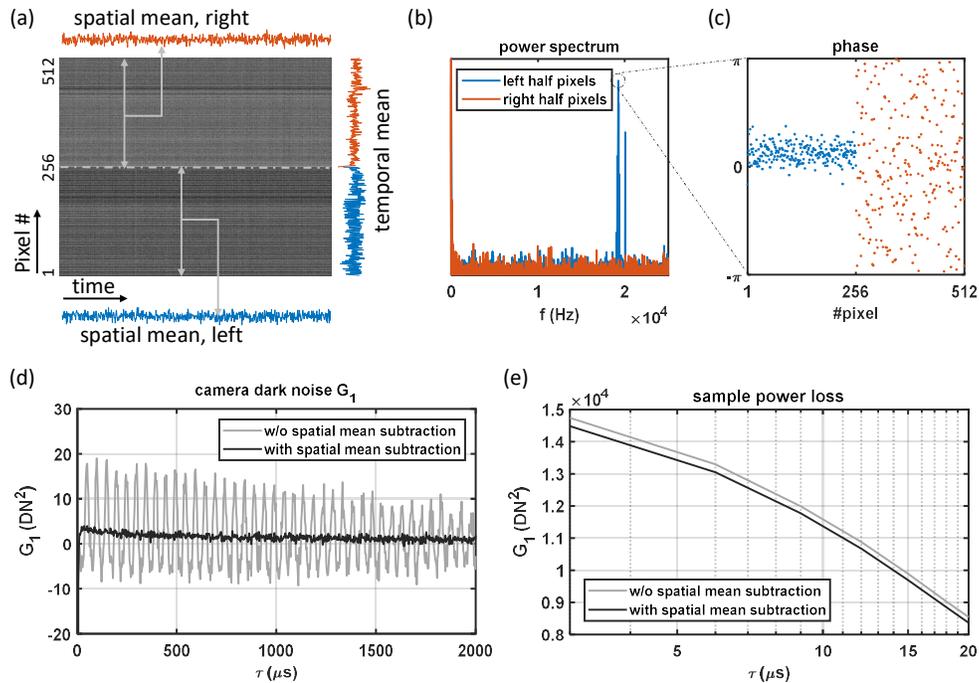

**Fig. 7** Removing spatially correlated camera oscillations by spatial mean subtraction. (a) Adding spatial mean subtraction. (b) and (c) Camera oscillation characterization. (d) Spatial mean subtraction removes the noise in $G_1$ introduced from the camera oscillation. (e) Spatial mean subtraction modestly affects the iDWS signal.

### Portable system

*Rationale:* One aim in this work is to engineer a portable iDWS system for space-starved clinical settings where large, pneumatically isolated optical tables are impractical. We hypothesized that without proper isolation, environmental vibrations can alter detected reference light, leading to fluctuations that overwhelm small sample signals. Thus we aimed to characterize the role of such factors in limiting long S-C separation iDWS measurements, and critically, to minimize them through system design.

*Experimental Approach:* We built a cart-based, portable iDWS system (lateral footprint: 1.5 feet x 2 feet, **Fig. 8a**). We assessed system stability through measuring S out C out reference $G_1$ (**Fig. 9**) under the conditions shown in **Table 1**. Besides passive air isolation of the interferometer (**Fig. 8b** yellow arrow), which is most crucial (**Fig. 9a**), the following features were found to aid system stability and overall performance:

1. Retractable casters with rubber feet (**Fig. 8b** inset, **Fig. 9b**), which support the cart as casters retract during the measurement.
2. Interferometer on the lowest shelf of the cart (**Fig. 8**b, **Fig. 9c**).
3. Sorbothane feet under the interferometer breadboard (**Fig. 8**b white arrow, **Fig. 9**d).
4. Camera height in good vertical alignment with the reference light (**Fig. 1** vertical view, **Fig. 9**e), which should make the system less sensitive to vertical vibrations.
5. Camera pixel vertical binning, which increases effective pixel height from 10 μm to 20 μm. (**Fig. 1** *en face*, **Fig. 9**f),

which will make the system less sensitive to vertical vibrations.

In summary, the portable system measurements incorporating aforementioned features is indistinguishable from those made with the interferometer on a pneumatically isolated optical table (**Fig. 9**a), marking an advance towards clinical neuromonitoring with interferometric diffuse optics.

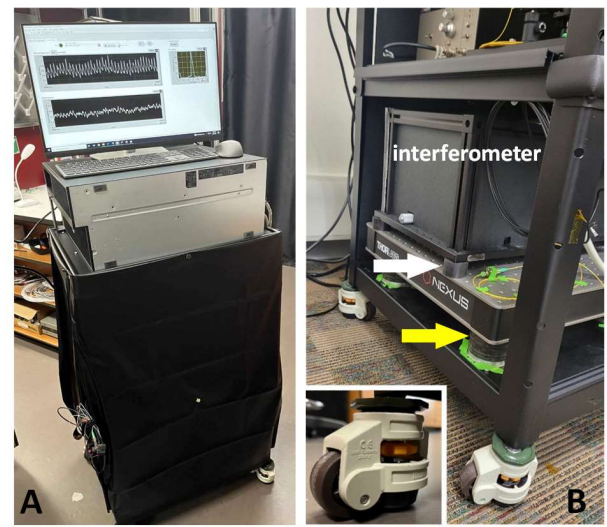

**Fig. 8** (a) Cart-based iDWS system enables portable CBFi monitoring. Retractable casters (b inset) and passive air isolation for the interferometer (b, yellow arrow), which is placed on the bottom shelf, help to stabilize the iDWS system.

**Table 2.** Combinations of design features reported in **Fig. 9**.

|  | Passive air isolation for interferometer | Rubber feet (casters retracted) | Interferometer on low shelf | Interferometer on sorbothane feet | Camera height aligned | Vertical pixel binning |
|---|---|---|---|---|---|---|
| Fig. 9a | y vs. n | y | y | y | y | y |
| Fig. 9b | n | y vs. n | y | y | y | y |
| Fig. 9c | n | y | y vs. n | y | y | y |
| Fig. 9d | n | y | y | y vs. n | y | y |
| Fig. 9e | n | y | y | y | y vs. n | y |
| Fig. 9f | n | y | y | y | y | y vs. n |

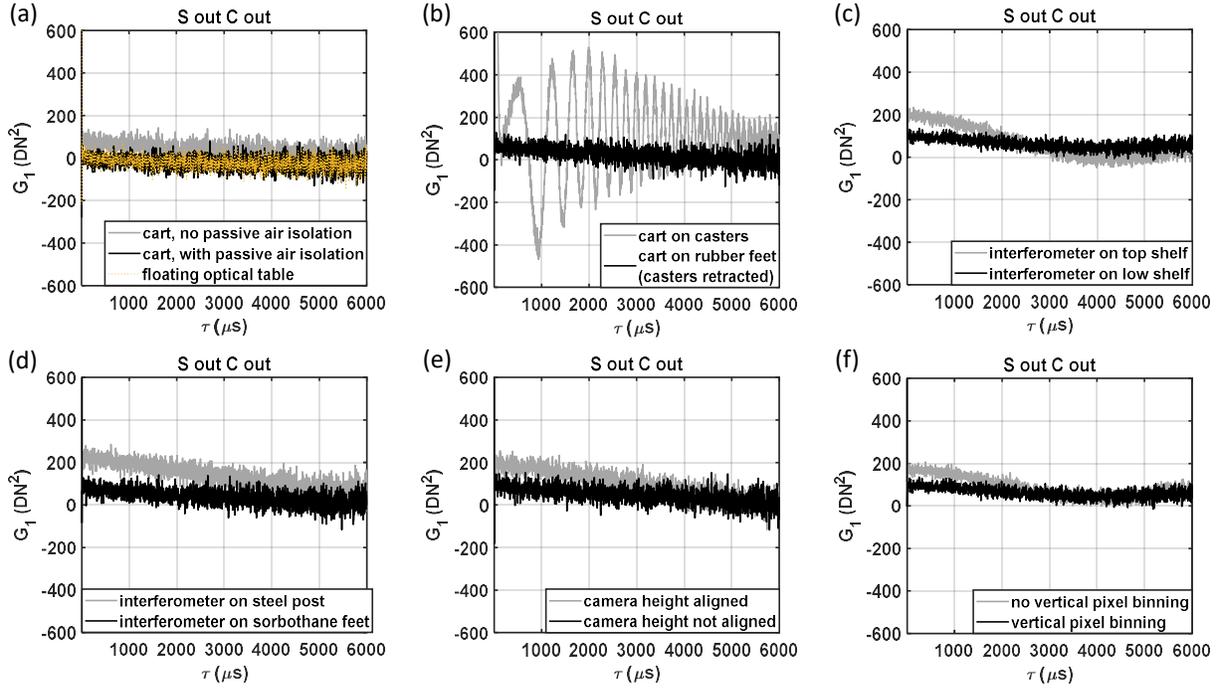

**Fig. 9** (a) Benchmark S out C out reference $G_1$ shows effectiveness of isolation strategy, which achieves comparable rejection of spurious signals to a floating laboratory optical table. Individual effect of each approach on stability the portable iDWS system: (b) rubber feet (c) lower shelf (d) sorbothane feet (e) alignment of camera height (f) tall camera pixel. Note that even a spurious $G_1$ amplitude of ~100 $DN^2$ is considered significant for S-C separation measurements (**Fig. 14**a).

### *Increasing full well capacity (FWC)*

*Rationale:* Increasing FWC to increase reference photoelectron counts can move the performance closer to the shot noise limit, a regime where reference shot noise dominates all other noise sources and the SANR becomes independent of reference counts [17]. Overall, performance can be improved, provided that other adverse effects of the increased FWC, including reduced heterodyne signal bit depth and increased contribution to $G_1$ from the spurious paths, are manageable.

*Experimental Approach:* We tested the camera performance at three different FWC levels: 22.7, 15.1 and 10.1 ke- (**Fig. 10**). We measured shot noise fraction [17] and $g_1$ noise variance for each case. As expected, we observed higher shot noise fraction at higher FWC level (**Fig. 10**), and accordingly the $g_1$ noise variance decreased with increasing FWC as expected (**Table 3**).

Adverse effects of increasing FWC must be mentioned. First, increasing FWC will reduce heterodyne signal bit depth. For the particular camera we used in this study (spL4096-140km, Basler), the highest FWC will reduce bit depth by less than 1 bit, which was not found to be significant. Second, increasing FWC will also require increasing the reference power to fill the camera to the same grey level and to gain the benefit of moving the system further into the shot-noise-limited regime. With increased reference power, contribution from the spurious paths to $G_1$ (**Fig. 6**a), will increase relative to the desired heterodyne signal. To attenuate the contribution of these spurious paths, we employed a coherence filtering approach as described in ***Coherent filtering of spurious paths***.

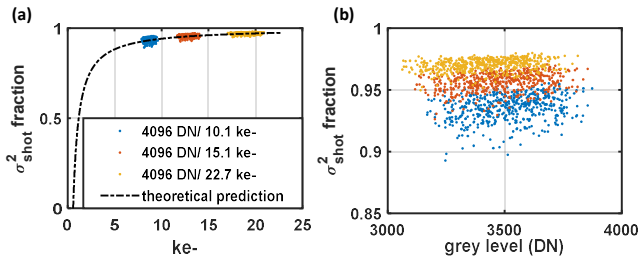

**Fig. 10** Operating at higher FWC moves the performance closer to the shot noise limit. (a-b) Fraction of noise variance attributable to reference shot noise increases with higher FWC,

**Table 3.** Comparison of $g_1$ noise variance at different FWCs.

| FWC (ke-) | $g_1$ noise variance* |
|---|---|
| 22.7 | $0.0104 \pm 0.00036$ |
| 15.1** | $0.0108 \pm 0.00040$ |
| 10.1 | $0.0109 \pm 0.00043$ |

\* $g_1$ noise variance averaged from $\tau = 0.45$ to 6 ms (n = 150 to 2000).
\** camera default setting

### Increasing camera duty cycle

*Rationale:* Camera dead time required for readout reduces the SANR by reducing the number of sample photons observed by the sensor. We aim to minimize this unnecessary loss in SANR by moving to a longer exposure time. Admittedly, sampling of the autocorrelation will be reduced and decorrelation during the exposure time will be increased [17, 24]. Both of these effects could adversely affect BFI recovery and must be considered while employing this strategy.

*Experimental Approach:* Autocorrelation SNR increases quadratically with camera duty cycle, for a fixed integration time, at a fixed sample rate. As the camera (spL4096-140km, Basler) has a fixed dead time (1.9 µs), we could increase duty cycle by increasing the line period, thus sacrificing sampling rate (**Fig. 11**a). In this case we partially compensate the loss in sampling rate by an accurate estimate of the zero lag of the autocorrelation (see **Cross pixel correlation for accurate estimation of autocorrelation at zero time lag**).

We first validated the increase in SANR by increasing duty cycle in phantom tests. We acquired data at duty cycles of 37%, 53% and 59% by setting the line period to 3 µs, 4 µs and 4.6 µs respectively, with reference arm power adjusted in each case to achieve roughly the same count level. Data were acquired at camera default FWC of 15.1 ke-. SANR was found to increase proportional to duty cycle and exposure time (**Fig. 11**b). Note the increase in SANR is a direct result of increasing exposure time, and detecting more sample photoelectrons.

We then validated the effect of increasing duty cycle in *in vivo* human forehead measurements at 3.5 cm S-C separation. Data were acquired at a duty cycle of 37% and 59% respectively, at a FWC of 15.1 ke-. The rBFi time courses (0.05 s integration time, 100 Hz sampling rate) show overall improvement in the recovered rCBFi waveform at the longer duty cycle (**Fig. 11**c-d), in spite of the reduced sampling rate.

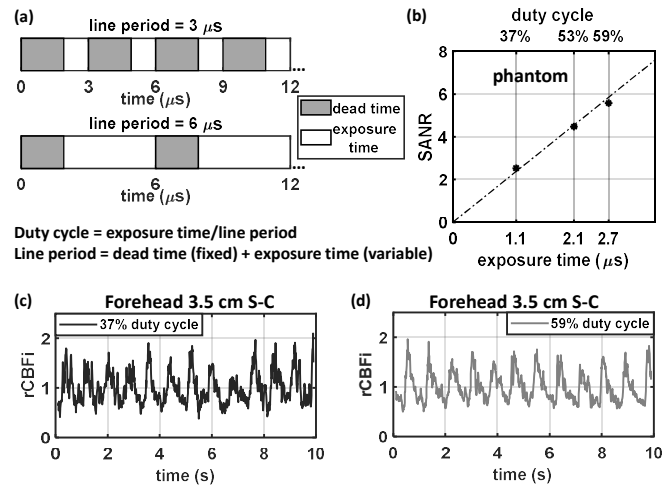

**Fig. 11** (a-b) Increasing camera duty cycle by increasing exposure time and line period improves SANR. In spite of the reduced sampling, rCBFi time courses are improved (c-d).

### MMF illumination

*Rationale:* Single-mode fiber (SMF) coupling can be sensitive to environmental vibrations and high SMF coupling efficiencies >70% can be challenging. Moreover, single mode illumination presents a possible retinal hazard that must be addressed through engineering controls.

*Experimental Approach:* The initial iDWS design coupled light to SMF, with a fused fiber coupler to split light between reference and sample arms (**Fig. 1**). In the new design, the order of splitting and fiber coupling is reversed. Thus, light is independently coupled to an SMF in the reference arm, which can tolerate losses, and an MMF, which affords high efficiency, in the sample arm (**Fig. 12**). This approach enables higher illumination power, thus higher SANR, with the same laser output power and is more eye-safe than SMF illumination.

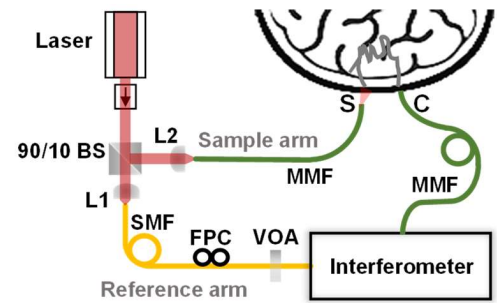

**Fig. 12.** MMF illumination increases laser coupling efficiency and etendue compared to SMF illumination, boosting illumination power and SANR, while improving light safety. BS: beamsplitter.

Experimentally, the MMF [fiber: Thorlabs GIF625, fiber collimator (L2 in **Fig. 12**): Thorlabs F110-APC-850] enabled a ~50% increase in free space coupling, compared to the SMF [fiber: Thorlabs 780HP, focusing lens (L1 in **Fig. 12**): Thorlabs A375TM-B]. Together with an upgrade of the laser, we were able to increase the sample illumination power from up to 70 mW (Vescent Photonics D2-100-DBR-852-HP1 [17, 21, 25, 26]) to up to 140 mW (Vescent Photonics D2-200-DBR-852-HP) in this work.

### Cross pixel correlation for accurate estimation of autocorrelation at zero time lag

*Rationale:* In iDWS, an accurate estimate of $G_{1,binned}$ at zero temporal lag can reveal tissue absorption changes [17], and aid measuring BFI at depth [17, 21, 25, 26], where autocorrelation decay rate is fast and sampling is limited. Especially if, as in this work, line rate is sacrificed to improve duty cycle (see **Increasing camera duty cycle**), an accurate estimate of the zero temporal lag can help compensate the loss in autocorrelation sampling rate. However, the autocorrelation at zero temporal lag is contaminated by a bias due to shot noise from the reference arm (**Fig. 2**b) [17].

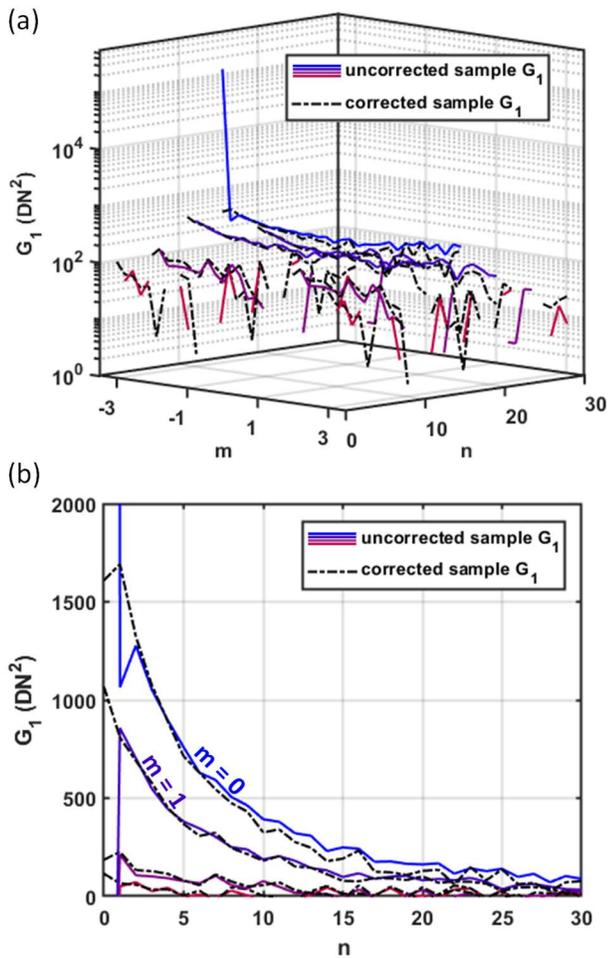

**Fig. 13** Three-dimensional (a) and two-dimensional (b) views of the spatiotemporal autocorrelation, $G_1$. $G_1(m\neq0, n=0)$, as defined in **Eq. (1)**, is free of shot noise bias, and thus can help to estimate of $G_1(m=0, n=0)$.

Our zero temporal lag estimation procedure is based on the principle that shot noise is uncorrelated between different populations of photoelectrons. As a result, shot noise bias appears in the autocorrelation $G_1$ only at zero spatiotemporal lag, $m=0$ and $n=0$ [**Eq. (1)**], as shown in **Fig. 13**a. Previously we used a fixed pattern noise (S out C in $G_1$) subtraction technique to correct the shot noise at $G_1(m=0, n=0)$ [17]. However, for low SANR data, small discrepancies in shot noise between full signal acquisition and S out C in acquisition will lead to errors at zero temporal lag (**Fig. 13**b dotted line, m=0). We observe that a more accurate full temporal autocorrelation can be obtained at non-zero spatial lags (e.g. **Fig. 13**b, dotted line m=1). Practically, we apply pixel binning to increase SANR [17], so we must extend this idea to $G_{1,binned}$. Basically then, the idea is to use spatial cross-correlations [i.e. $G_1(m\neq0, n)$] to estimate the zero temporal lag in $G_{1,binned}$. In doing this, we assume that $G_1(m=0, n)$ and $G_1(m\neq0, n)$ have the same decay rate along time lag $\tau$.

The optimally binned autocorrelation, as previously presented [17], consists of two terms; the first is the sum of zero spatial lag components, and the second is the sum of non-zero spatial lag components:

$$G_{1,binned}(n) = \underbrace{\sum_i \sum_p H_{p,p} s(p,i) s(p,i-n)}_{\text{zero spatial lag}}$$
$$+ \underbrace{\sum_i \sum_{p\neq q} H_{p,q} s(p,i) s(q,i-n)}_{\text{non-zero spatial lag}} \quad (6)$$

To correct $G_{1,binned}(n=0)$ for the shot noise bias, we calculate the second term and the sum of terms in **Eq. (6)**. Next we do a regression between the non-zero temporal lag points of the two to get a scaling factor. Then we apply the scaling factor to the zero temporal lag point of the second term and take it as the $G_{1,binned}(n=0)$.

### Overall testing and validation

Combining all of the aforementioned iDWS improvements (**Table 4**) and post processing methods, we took measurements on an adult subject's forehead (Fitzpatrick 4) at 4 and 4.5 cm S-C separations. Unnormalized autocorrelations (**Fig. 14**a), integrated over one heartbeat, with shaded standard deviation over 20 heartbeats, showed comparable noise levels, with ~47% reduction in signal at 4.5 cm compared to 4 cm; while normalized autocorrelations (**Fig. 14**b) showed higher relative noise levels at 4.5 cm. CBFi time course at 0.125 s integration time and 100 Hz sampling (recovered by DCS semi-infinite model fitting [27]) for both separations show clear pulsatile waveforms (**Fig. 14**c-d).

**Table 4.** Contributions of individual system parameters to total $SNR_{ac}$ improvement.

| Parameters | Ref. [17] | Current work | $SNR_{ac}$ improvement |
|---|---|---|---|
| $N_{channel}$ | 192 | 338 | 1.76× |
| FWC (ke-) | 22.7 | 15.1 | 1.04× |
| Duty cycle | 37% | 68% | 3.38× |
| Illumination power (mW) | 70 | 140 | 4× |
| **Total** | | | **24.75×** |

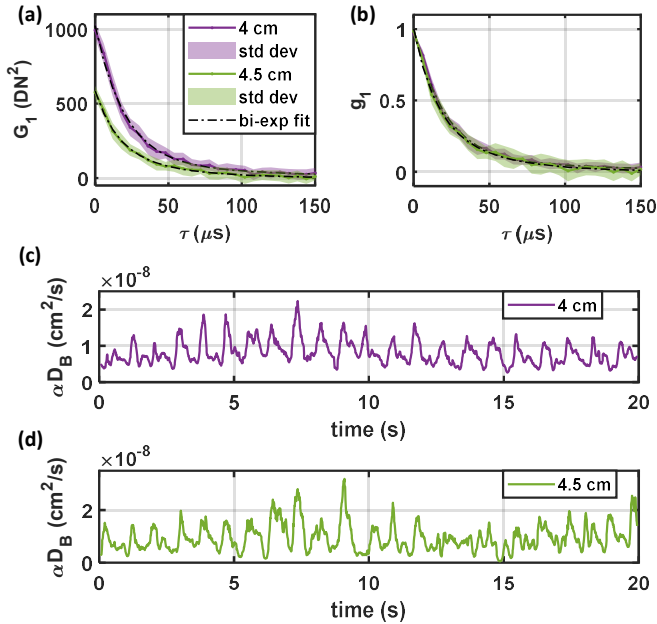

**Fig. 14**. CBFi measurements at 4 and 4.5 cm S-C separations. (a) $G_1$ (heartbeat integrated, ± standard deviation). (b) $g_1$ (heartbeat integrated, ± standard deviation). (c)-(d) CBFi time courses, with 0.125 s integration time, and 100 Hz temporal sampling.

### Preliminary clinical tests

The portable iDWS system was wheeled into the NYU Langone Hospital-Brooklyn Neuro ICU (**Fig. 15**a), where a female with Moyamoya disease (36 years old, Fitzpatrick 3) was recruited while undergoing continuous video electroencephalography post-seizure. Informed consent was obtained from subject's legally authorized representative. Experimental procedures were approved by NYU Langone IRB. Measurements were performed on the forehead at 3.5 and 4 cm S-C separations (**Fig. 15**b). CBFi time courses, recovered by a DCS semi-infinite model [16], with 0.125 s integration time and sampled at 100 Hz, showed clear pulsatile waveforms. Due to limited space for applying and fixing the optical probe given that the subject was wearing electroencephalogram electrodes on her head, we only attempted to acquire data at a maximum of 4 cm S-C separation. We expect to increase S-C separation in patients to 4.5 cm by improving the probe design and probe-to-tissue coupling.

### DISCUSSION

In this work we describe optimization of a continuous wave (CW) iDWS system. The prior iteration of the system achieved CW measurements of pulsatile CBFi with a 0.1 s integration time at 3.5-4.0 cm S-C separation [17]. With careful optimization of multiple parameters, we are able to observe pulsatile CBFi with 0.125 s integration time from adult forehead at 4.0-4.5 cm S-C separation. The engineering of a demonstrably stable cart-based iDWS system represents a milestone in clinical translation of interferometric diffuse optical methods.

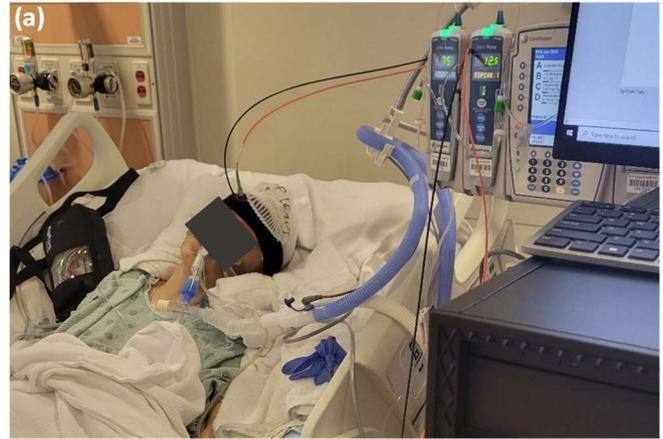

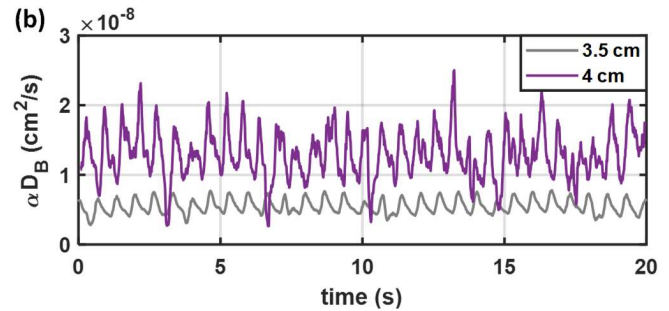

**Fig. 15**. CBFi measurements in Neuro ICU with the portable iDWS system. (a) Bed side measurement setup. (b) Pulsatile CBFi at 3.5 and 4 cm S-C separations (0.125 s integration time, 100 Hz sampling).

It is instructive to compare our approach broadly with other members in the growing family of DCS-inspired approaches that exploit parallelism to measure deep CBFi [28]. A promising recent advance in autocorrelation sampling techniques is SPAD array DCS [13], which achieved overall performance comparable to this work at a similar wavelength. Since iDWS samples the field autocorrelation $G_1$, our 6 µs sampling is equivalent to ~3 µs sampling of the intensity autocorrelation (assuming exponential decay) [13]. However, the presented iDWS approach has several advantages. First, DCS inherently measures intensity not field. Intensity does not incorporate phase information, and requires the Siegert relationship to recover the field autocorrelation where a coherence parameter must be independently calibrated or assumed. Second, the abilities to sample CBFi more frequently than the integration time and to change integration time in post-processing are not afforded by current SPAD array implementations which save only selected autocorrelation lags [29]. Finally, a major advantage of iDWS over SPAD array DCS is cost. We estimate that our CMOS sensor is approximately two orders-of-magnitude less expensive than a 512x512 SPAD array. We acknowledge that the cost of SPAD arrays may decrease as more applications emerge. On the other hand, the cost of fast CMOS sensors is likely to decrease as well.

Comparisons with other CBFi technologies are less straightforward. For instance, speckle contrast optical spectroscopy (SCOS) can achieve very large source-collector separations [30-32], and at first glance, the performance may seem superior to this work. However, the exposure times required in SCOS imply a considerable loss in bran sensitivity and increase in scalp sensitivity compared to the short exposures used here [33]. Comparisons with 1064 nm interferometric methods must take into account the inherent physical advantages of the 1064 nm wavelength, and the ~5x higher cost of InGaAs sensors [18]. We do expect that a 1064 nm iDWS system, optimized using the concepts presented here, should easily achieve pulsatile CBFi at >5 cm S-C separation.